\documentclass[prb,aps,twocolumn,superscriptaddress,floatfix,citeautoscript]{revtex4-2}
\usepackage{graphicx,rotating,subfigure,amsmath,amsfonts,amssymb,delarray,color}
\usepackage{hyperref}
\usepackage{xcolor}
\usepackage{soul}
\usepackage{dsfont}
\usepackage[T1]{fontenc}
\usepackage{physics}
\usepackage{multirow}
\usepackage{comment}

\def\12{\frac{1}{2}}

\begin{document}
%\bibliographystyle{apsrev}

%%%%%%%%%%%%%%%%%%%%%%%%%%%%%%%%%%%%%%%%%%
%%%%%%%%%%%%%%%%%%%%%%%%%%%%%%%%%%%%%%%%%%

\title{Absence of true localization in many-body localized phases}

%%%%%%%%%%%%%%%%%%%%%%%%%%%%%%%%%%%%%%%%%%
%%%%%%%%%%%%%%%%%%%%%%%%%%%%%%%%%%%%%%%%%%

\author{Maximilian Kiefer-Emmanouilidis}
\affiliation{Department of Physics and Research Center OPTIMAS, University Kaiserslautern, 67663 Kaiserslautern, Germany}
\affiliation{Department of Physics and Astronomy, University of Manitoba, Winnipeg R3T 2N2, Canada}
\author{Razmik Unanyan}
\affiliation{Department of Physics and Research Center OPTIMAS, University Kaiserslautern, 67663 Kaiserslautern, Germany}
\author{Michael Fleischhauer}
\affiliation{Department of Physics and Research Center OPTIMAS, University Kaiserslautern, 67663 Kaiserslautern, Germany}
\author{Jesko Sirker}
\affiliation{Department of Physics and Astronomy, University of Manitoba, Winnipeg R3T 2N2, Canada}
\affiliation{Manitoba Quantum Institute, University of Manitoba, Winnipeg R3T 2N2, Canada}

\date{\today}

\begin{abstract}
We have recently shown that the logarithmic growth of the entanglement
entropy following a quantum quench in a many-body localized (MBL)
phase is accompanied by a slow growth of the number entropy,
$S_N\sim\ln\ln t$. 
%This violates the standard scenario of MBL and
%raises the question whether the observed behavior is transient or
%continues to hold at strong disorder in the thermodynamic limit.  
Here
we provide an in-depth numerical study of $S_N(t)$ for the disordered
Heisenberg chain 
%and find strong evidence that this behavior persists
%even at strong disorder. 
and show that this behavior is not transient and persists even for
very strong disorder. Calculating the 
%% MF 
truncated
R\'enyi number entropy
%% JS
$S_N^{(\alpha)}(t)=(1-\alpha)^{-1}\ln\sum_n p^\alpha(n)$ for
$\alpha\ll 1$ and $p(n)>p_c$---which is sensitive to large number
fluctuations occurring with low probability---we demonstrate that the
particle number distribution $p(n)$ in one half of the system has a
continuously growing tail. This indicates a slow but steady increase
of the number of particles crossing between the partitions in the
interacting case, and is in sharp contrast to Anderson localization,
for which we show that $S_N^{(\alpha\to 0)}(t)$ saturates for any
cutoff $p_c>0$. 
%% JS
We show, furthermore, that the growth of $S_N$ is {\it not} the
consequence of rare states or rare regions but rather represents
typical behavior. These findings provide strong evidence that the
interacting system is never fully localized even for very strong but
finite disorder.
\end{abstract}

\maketitle

%%%%%%%%%%%%%%%%%%%%%%%%%%%%%%%%%%%%%%%%%%
%%%%%%%%%%%%%%%%%%%%%%%%%%%%%%%%%%%%%%%%%%
\section{Introduction}
%%%%%%%%%%%%%%%%%%%%%%%%%%%%%%%%%%%%%%%%%%
%%%%%%%%%%%%%%%%%%%%%%%%%%%%%%%%%%%%%%%%%%

In a one-dimensional system of free particles with short-range
hoppings, even the smallest amount of potential disorder leads to a
localization of the single particle wave functions, a phenomena termed
Anderson localization
\cite{Anderson58,AbrahamsAnderson,AndersonLocalization}. 
A question which has remained open for more than 50 years is whether
or not localization is also possible in an interacting many-body
system. This question has been put back to the forefront of research
in condensed matter physics by a seminal work by Basko, Aleiner, and
Altshuler arguing perturbatively that at weak interactions a
metal-insulator transition, i.e.~a many-body localization (MBL)
transition, will occur at some finite temperature
$T_c$ \cite{BaskoAleiner}. This work has sparked a number of studies of
possible ergodic-MBL transitions in disordered lattice models. The
most studied of these models is the spin-1/2 Heisenberg chain with
local magnetic fields drawn from a box distribution
\cite{OganesyanHuse,PalHuse,Luitz1,Luitz2,NandkishoreHuse,AltmanVoskReview,SerbynPapic,BarLevCohen,VoskHusePRX,PotterVasseurPRX,VoskAltman,VoskAltman2,PietracaprinaMace}, which is equivalent 
to the fermionic t-V model with potential disorder.
% by Jordan-Wigner transform. 
The results have been interpreted in terms of an
ergodic-MBL transition at finite disorder strength. Under the
assumption of limited level attraction, perturbative arguments for the
stability of an MBL phase in spin chains have been put forward
\cite{RosMuellerScardicchio,Imbrie2016} but a rigorous proof is lacking. 
Very recently, numerical studies have cast some doubt on the stability
of MBL in the thermodynamic limit
\cite{SuntajsBonca,SuntajsBonca2,Znidaric2018,SelsPolkovnikov}. However, the
interpretation of these results is still a matter of debate \cite{Abaninrecent}.
% with several groups arguing that the system sizes which can be studied by
% exact diagonalization are too small to perform a scaling
% analysis
 %,Localisationconference}.

\begin{figure}
	\includegraphics*[width=0.9\columnwidth]{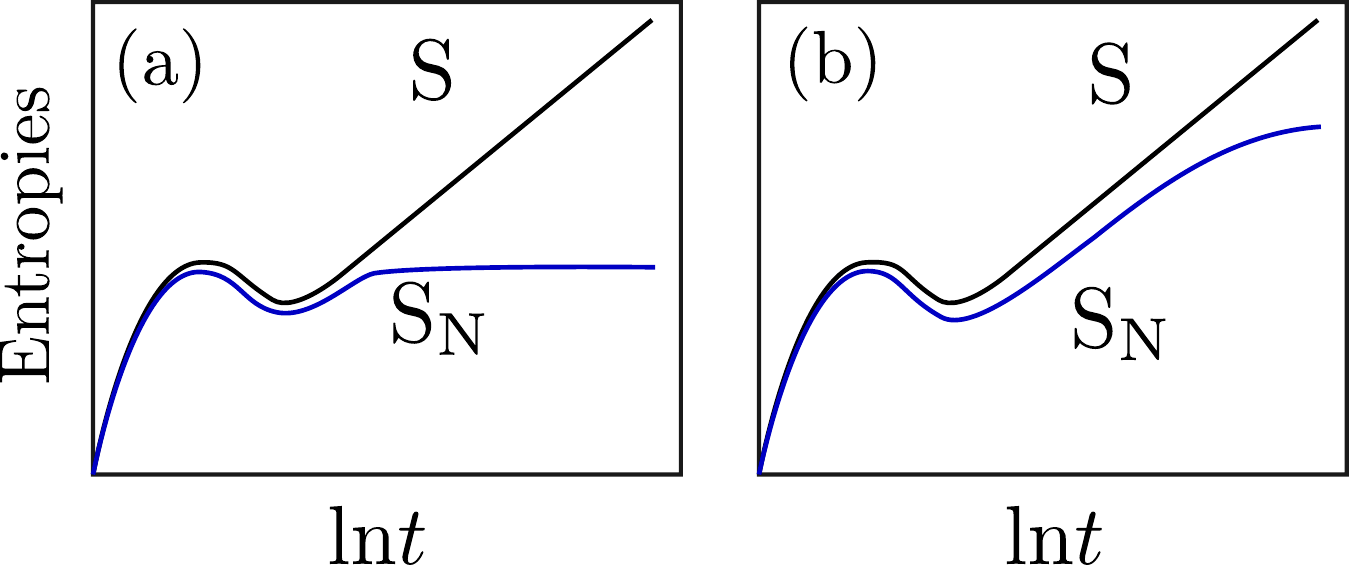}
	\caption{(a) Standard MBL scenario: The number entropy
	saturates. A further logarithmic increase of the entanglement
	entropy is caused entirely by the configurational entropy. (b)
	Alternative scenario: The number entropy never saturates. The
	logarithmic increase of the total entanglement coexists with
	an $S_N\sim\ln\ln t$ increase of the number entropy.}
\label{Fig1f}
\end{figure}

Another recent development is the study of symmetry-resolved
entanglement measures
\cite{WisemanVaccaro,Rakovszky2019,LukinRispoli,Bonsignori2019,MurcianodiGiulio,MurcianodiGiulio2}. 
For a system with particle number conservation, 
% (or total spin conservation in the spin language) 
the von-Neumann entanglement entropy $S$ can be split into two
contributions
\begin{eqnarray}
\label{SvN}
S &=& S_N + S_c\, ,\quad S_N = -\sum p(n)\ln p(n)\, , \nonumber \\
S_c &=& -\sum_n p(n)\tr[\rho(n)\ln\rho(n)]. 
\end{eqnarray}
Here $S_N$ is the number entropy which is entirely characterized by 
the probability  $p(n)$ to find $n$ particles in the
considered subsystem. $S_c$ is the configurational
entropy with $\rho(n)$ being the block of the reduced density
matrix with particle number $n$. While using symmetries is, on the one
hand, of fundamental interest from a quantum information perspective
to calculate the amount of operational entanglement which is available
\cite{WisemanVaccaro,SchuchVerstraeteCirac,SchuchVerstraeteCirac2,MonkmanSirker}, it is, on the 
other hand, also helpful to understand how much of the entanglement is
caused by particle fluctuations and how much is due to the
superposition of different configurations in a sector of constant
particle number. The usefulness of this approach
% to investigate many-body localization 
has recently been demonstrated in a cold-atomic gas experiment where
entanglement following a quench in a one-dimensional Aubry-Andr\'e
Bose-Hubbard model was studied
\cite{LukinRispoli}. The experimental results have been interpreted in
terms of a number entropy which saturates and a configurational
entropy which then continues growing logarithmically on top of the
constant number entropy (see Fig.\ref{Fig1f}a).
%
%%%%%%%%%%%%%%%%%%%%%%%%%%%%%%%%%%%%%%%%%%
%%%%%%%%%%%%%%%%%%%%%%%%%%%%%%%%%%%%%%%%%%
%
%
%This is consistent with the standard picture
%of MBL phases according to which the dynamics is described by an
%effective model of exponentially many conserved charges $[H,\eta_l]=0$
%with
%
%\begin{equation}
%\label{Heff}
%H = \sum_l \varepsilon_l \eta_l + \sum_{l,l'} J_{ll'} \eta_l \eta_{l'} + \cdots \, .
%\end{equation}
%
%Here $\varepsilon_l$ are random energies and $J_{ll'}$ are coupling
%constants which are exponentially decaying with distance between the
%sites $l$ and $l'$. The effective model \eqref{Heff} describes a phase
%with no particle transport while different configurations get
%entangled over time $t\sim \e^\ell$ leading to $S\sim S_c\sim \ln
%t$. 
%
The resulting logarithmic growth of the total entanglement entropy has been
confirmed in several numerical studies
\cite{ZnidaricProsen,BardarsonPollmann,AndraschkoEnssSirker,EnssAndraschkoSirker}. 
The behavior of the number entropy $S_N$, however, has received much
less attention. Very recently, we have shown that in the numerically
accessible time regime the logarithmic growth of entanglement in the
MBL phase is accompanied by a growth $S_N\sim\ln\ln t$ of the number
entropy (see Fig.\ref{Fig1f}b) \cite{KieferUnanyan1,KieferUnanyan2}. If
this behavior does persist in the thermodynamic limit for all finite
disorder strengths then the MBL phase would 
%% JS
ultimately not be localized and the system would always remain
ergodic.

The purpose of this paper is to further study the two scenarios for
the entanglement evolution in MBL phases, shown schematically in
Fig.\ref{Fig1f}. To do so, we will carefully study the timescales
where the scaling behavior holds as well as the distributions of the
total entanglement entropy and of the number entropy. It has also been
suggested recently by Bar Lev and Luitz \cite{LuitzBarLev} that the
increase of the number entropy observed in our previous publication,
Ref.~\cite{KieferUnanyan2}, might be a result of disorder
strengths that were still relatively close to the transition point. In
order to address this point, we will extend our numerical study to
disorder strengths up to twice of what is believed to be the critical
value.
%% JS
It is known that on the localized side but still close to the
ergodic-MBL transition, rare regions with less disorder can cause a
very slow dynamics \cite{Gopalakrishnan2015,AgarwalAltman} and can
destabilize the MBL phase in small systems. In order to exclude such a
scenario we will compare the average with the median number entropy
and show that the observed growth of the number entropy is {\it not} a
consequence of rare initial states or rare regions.

To further investigate if the observed slow growth of the number
entropy is transient, we study the time evolution of the (discrete) probability
distribution $p(n,t)$. If MBL is associated with a very slow formation
of localized states, there could be a long transient time period where
probabilities redistribute in a very narrow range of particle numbers,
while larger fluctuations are strictly suppressed. The number entropy
is not sufficiently sensitive to large particle number fluctuations occuring
with small probability and thus cannot 
unambiguously exclude such a
scenario. A much more sensitive measure are the number R\'enyi entropies
\begin{equation}
S_N^{(\alpha)} = (1-\alpha)^{-1} \ln \, \sum_{n=0}^\infty p^\alpha(n),\label{SN_alpha}
\end{equation}
with $\alpha \ll 1$.
%% JS
% and $p(n) > p_c$. 
The family of R\'enyi entropies provides information
about different characteristics of the probability
distribution. $S_N^{(1)}= -\sum_n p(n) \ln p(n)$, for example, is the
well-known Shannon entropy. For growing values of $\alpha$ the R\'enyi
entropies are increasingly determined by the largest probability
values.  $S_N^{(\infty)}= - \ln p_\textrm{max}(n)$, in particular, is
given by the logarithm of the maximum probability. For decreasing
values of $\alpha \ll 1$, $S_N^{(\alpha)}$ becomes increasingly
sensitive to all non-vanishing probabilities including those that are
small. 
Taking the limit $\alpha\to 0$
%% JS
% with $p(n)>p_c$, 
$S_N^{(\alpha)}$ gives the so-called Hartley number entropy, which
essentially counts all values of $n$ which have a non-vanishing
probability $p(n)$.
%% JS
% larger than some cutoff $p_c$. 
In order for the Hartley entropy to become a useful physical quantity
to investigate the properties of $p(n)$ one has to introduce a cutoff
probability $p_c>0$.  E.g., if $p(n)> p_c$ for $M<N_s$ values of $n$,
then $S_N^{(0)}=\ln M$.
%Introducing a cut-off $p_c$ by setting
%setting all probabilities $p(n)< p_c$ to zero and renormalizing the
%distribution, the Hartley entropy gives a measure for the number of
%values $n$ with probability larger than the cut-off.

%% JS
Our paper is organized as follows: In Sec.~\ref{Model} we introduce
the model and notation and also discuss the numerical methods and the
averaging procedure. In Sec.~\ref{Results} we then present the results
of our numerical investigations for the entanglement and number
entropy. The section is subdivided into two subsections, dealing with
the coexistence of the growth of $S$ and $S_N$ and a comparison
between the average and the median, and the distributions of
entanglement for different realizations, respectively. The results for
the Hartley number entropy are discussed in Sec.~\ref{Sec_Hartley}. In
Sec.~\ref{Concl} we present our conclusions and discuss some of the
remaining open questions.

%%%%%%%%%%%%%%%%%%%%%%%%%%%%%%%%%%%%%%%%%%
%%%%%%%%%%%%%%%%%%%%%%%%%%%%%%%%%%%%%%%%%%
\section{Model and Methods}
\label{Model}
%%%%%%%%%%%%%%%%%%%%%%%%%%%%%%%%%%%%%%%%%%
%%%%%%%%%%%%%%%%%%%%%%%%%%%%%%%%%%%%%%%%%%

We concentrate here on the isotropic Heisenberg model in the fermionic
representation (t-V model)
\begin{equation}
\label{tV}
H = -J\sum_j \left\{(c^\dagger_j c_{j+1} +h.c.)+D_j n_j +V n_j n_{j+1}\right\} \, ,
\end{equation}
with nearest-neighbor interaction $V=2J$. We assume a half-filled
system and draw random values of the local potential from a box
distribution, $D_j\in [-D/2,D/2]$. Throughout, we are using open
boundary conditions. Note that in the notation used here, $D_j = 4h_j$
where $h_j$ are the local magnetic fields in the spin representation
used, for example, in Refs.~\cite{PalHuse,Luitz1,Luitz2}. We are
interested in the growth of entanglement following a quantum quench
from a random product state $|\Psi_0\rangle$. This state is
then time evolved, $|\Psi(t)\rangle = \exp(-iHt)|\Psi_0\rangle$. We
set $J=1$ throughout this paper.

For system sizes $L\leq 14$ we use exact diagonalizations of the
Hamiltonian matrix to obtain the time-evolved state
$|\Psi(t)\rangle$. We then calculate the reduced density matrix by
tracing out half of the system, $\rho = \tr_A
|\Psi(t)\rangle\langle\Psi(t)|$, and calculate the number distribution
$p(n,t)$. Typically, we pick $10,000$ random disorder configurations
and for each disorder configuration we average over $50$ random
half-filled initial product states. To avoid any possible issues due
to the double precision limitations of standard exact diagonalizations
\cite{ZhaoAndraschkoSirker}, we limit ourselves to system sizes where
the saturation times remain $\lesssim 10^{14}$.

As a complementary method, we use a Trotter-Suzuki decomposition of
the time evolution operator \cite{Suzuki1,Suzuki2,Trotter}. 
%Applying the decomposed operator onto the
%state, we can then obtain $\vert\Psi(t)\rangle$ without having to  deal
%with the full Hamiltonian matrix. 
This allows to reach larger system sizes; we restrict ourselves here
to $L\leq 24$---for even larger systems the computational cost of
calculating several thousand samples becomes prohibitive. Since the
Trotter error of the decomposition accumulates over time, the
simulation times for the chosen Trotter parameter $\delta t\sim
10^{-4}$ are limited to $t\lesssim 10^3$. Here, we typically average
over $1,500$ disorder realizations for $D\leq 28$ and $2,000$ for
$D>28$ and pick a random initial product state for each
realization. We note that the various entropies are calculated first
for each sample separately and are then, in a second step, 
%% JS
either averaged over all realizations or used to calculate the median.

%%%%%%%%%%%%%%%%%%%%%%%%%%%%%%%%%%%%%%%%%%
%%%%%%%%%%%%%%%%%%%%%%%%%%%%%%%%%%%%%%%%%%
\section
%\paragraph
{Entanglement entropy and number entropy}
\label{Results}
%%%%%%%%%%%%%%%%%%%%%%%%%%%%%%%%%%%%%%%%%%
%%%%%%%%%%%%%%%%%%%%%%%%%%%%%%%%%%%%%%%%%%

Here we present and analyze the results for the entanglement and
number entropies obtained by the numerical simulations described
above. We provide evidence that the unbounded growth of the number
entropy persists even deep in the MBL phase and that it is
associated with sub-exponential tails in the probability distribution
$p(S_N)$ of the number entropy, which grow in time. We also show that
not only the average number entropy grows as $S_N\sim\ln\ln t$ but
also the median number entropy, 
%%MF
providing evidence that the observed behavior 
is typical and not due to rare cases.

%% JS
%%%%%%%%%%%%%%%%%%%%%%%%%%%%%%%%%%%%%%%%%%
\subsection{Growth of $S(t)$ and $S_N(t)$}
\label{parallel}
%%%%%%%%%%%%%%%%%%%%%%%%%%%%%%%%%%%%%%%%%%

\begin{figure}
	\includegraphics*[width=0.9\columnwidth]{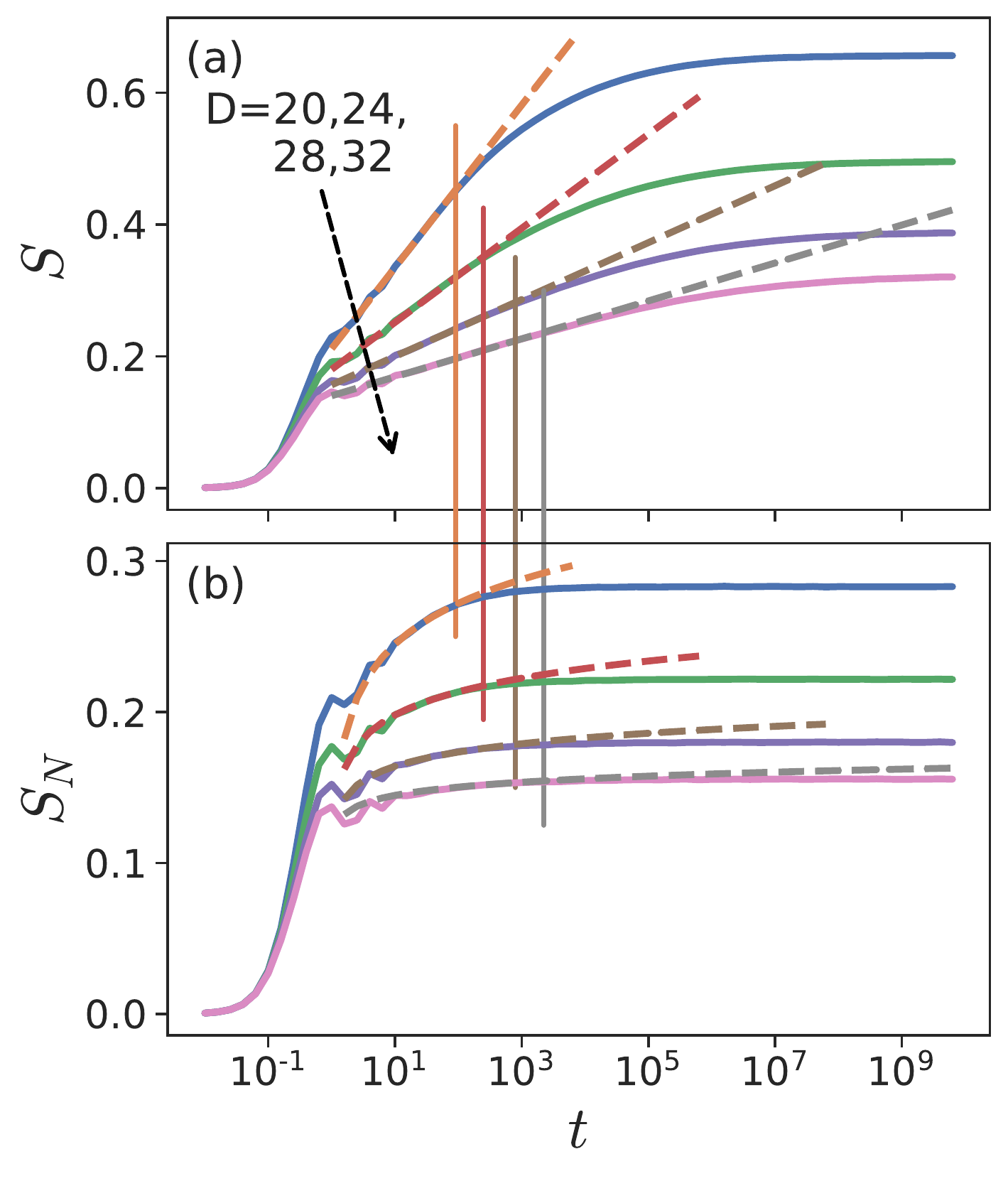}
	\caption{(a) Entanglement entropy, and (b) number entropy for
	$L=14$ and different disorder strengths $D>D_c$. The dashed
	lines are logarithmic (double logarithmic) fits,
	respectively. In all cases, finite-size saturation sets in at
	the same time scale (marked by vertical lines) in both quantities.}
\label{Fig2f}
\end{figure}

\begin{figure}
	\includegraphics*[width=0.9\columnwidth]{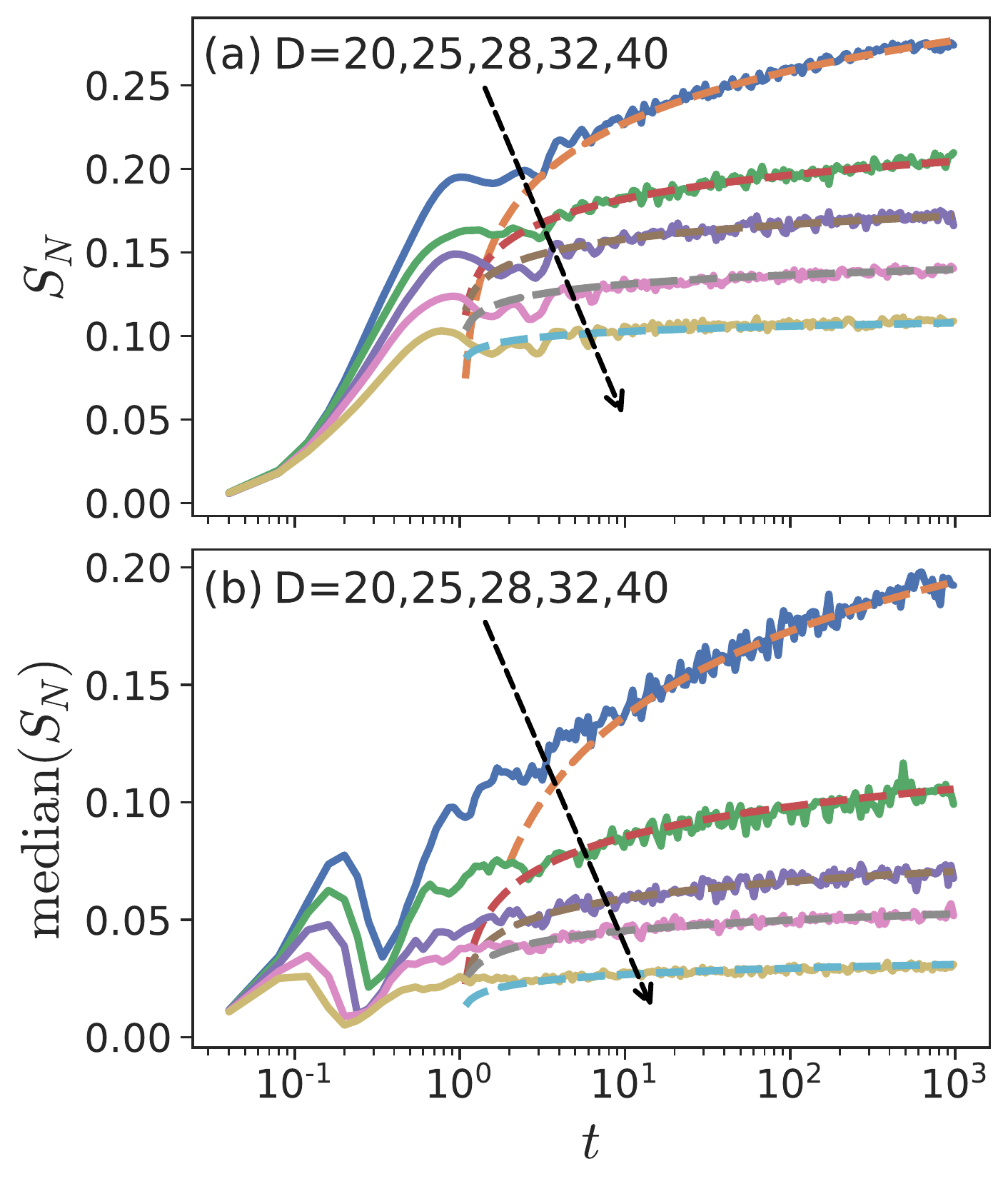}
	\caption{Averaged number entropy (panel (a)) and median number
	entropy (panel (b)) for $L=24$ and different disorder
	strengths $D>D_c$. The $S_N\sim\ln \ln t$ scaling (represented
	by the dashed-line fits) persists for all disorder strengths
	in {\it both} quantities up to the largest times reached in
	our simulations.}
\label{Fig3f}
\end{figure}

First, we want to demonstrate that the $S\sim \ln t$ growth of the
entanglement entropy and the $S_N\sim \ln\ln t$ growth of the number
entropy are intimately related and persist over the same time scales,
only limited by the considered system size. Second, we want to
demonstrate that this behavior is not restricted to a narrow range of
disorder strengths near the localization transition but is also
present deep in the MBL phase. Previous numerical calculations put the
critical disorder in the range $D_c \sim 14\dots 17$. In Fig.~\ref{Fig2f}, we
therefore present results for disorder strengths up to about twice the
critical value.
%
%%%%%%%%%%%%%%%%%%%%%%%%%%%%%%%%%%%%%%%%%%
%%%%%%%%%%%%%%%%%%%%%%%%%%%%%%%%%%%%%%%%%%
%
The main point we want to make is that both $S(t)$ and $S_N(t)$ start
to saturate due to the finite size of the system at {\it the same time
scale.} 
%We do observe this behavior for all system sizes and disorder
%strengths studied. Also for very large disorder 
We never find a case where $S_N$ starts to saturate while $S$ continues to
grow logarithmically as would be expected in the standard scenario,
Fig.~\ref{Fig1f}(a).
% Even for the largest
% system size which we have studied, $L=24$, 
We find, furthermore, that a perfect $S_N\sim\ln\ln t$ scaling holds
up to the largest simulation times even at very large disorder, see
Fig.~\ref{Fig3f}(a). Since there is still some debate about the
precise value of the critical disorder strength for the onset of MBL,
one could argue that our results in Ref.~\cite{KieferUnanyan2}
might only be valid close to the transition point
\cite{LuitzBarLev}. Our new results clearly show that this is not the
case.

%% JS
An important question then is what causes the slow growth of the
number entropy. In Ref.~\cite{AgarwalAltman} it has been argued that
rare thermal regions in the localized phase can dominate its
low-frequency response. To investigate whether or not the observed
growth is related to rare initial states or rare disorder
configurations leading to rare thermal inclusions, we show in
Fig.~\ref{Fig3f}(b) the median number entropy. 
%
%%%%%%%%%%%%%%%%%%%%%%%%%%%%%%%%%%%%%%%%%%
%%%%%%%%%%%%%%%%%%%%%%%%%%%%%%%%%%%%%%%%%%
%
This quantity is defined by sorting the number entropies for each
realization in terms of magnitude at a given point in time and then
choosing the value in the middle, for an odd number of realizations, or the average of the two middle values, for an even number of realizations. The answer is unambiguous: The median
number entropy shows the same double logarithmic growth in time as the
average number entropy. We conclude that the observed growth is not
the consequence of rare regions but rather represents the typical
behavior of the number entropy. The scenarios discussed in
Ref.~\cite{AgarwalAltman} do not explain our observations: We find
that the short time rather then the long time behavior is strongly
influenced by rare non-typical samples. When comparing
Fig.~\ref{Fig3f}(a) with Fig.~\ref{Fig3f}(b) we see that the main
qualitative difference is a supression of the initial approximatively
logarithmic increase in the median as compared to the average number
entropy. A natural explanation is that rare regions with little
disorder (called 'thermal regions' in Ref.~\cite{AgarwalAltman}) are
responsible. Indeed, we have shown that in the ergodic phase at small
disorder $S_N\sim\ln t$
\cite{KieferUnanyan2}. Excluding rare configurations from the average as we do when
calculating the median therefore strongly reduces the number entropy
at short times---and thus also the overall values for all times---but
does not change the scaling at long times.

%%%%%%%%%%%%%%%%%%%%%%%%%%%%%%%%%%%%%%%%%%
\subsection{Variance of entropies and entropy distribution functions}
%%%%%%%%%%%%%%%%%%%%%%%%%%%%%%%%%%%%%%%%%%

The numerical study of disordered systems requires a careful sampling
of disorder realizations. 
%% JS
So far, we have either considered averages over all realizations or
have calculated the median. As pointed out in Ref.~\cite{LuitzBarLev},
it is however useful to consider also the probability distributions of
these entropies w.r.t. the different realizations for a deeper
understanding of the underlying physics. In fact, numerical studies in
Ref.~\cite{LuitzBarLev} have shown that while the variance of the
asymptotic entanglement entropy $\Delta S$ approaches a constant value
with increasing system size,
%in the limit of strong disorder, 
the corresponding value of the number entropy, $\Delta S_N$, decreases.
% with system size.
% Furthermore these 
%numerical studies in Ref.~\cite{BarLevLuitz}
In addition, it was shown that the probability distribution $p(S)$ of
the entanglement entropy for a given system size $L$ and large
disorder has an exponential tail. In contrast, the probability
distribution of the corresponding number entropy $p(S_N)$ shows a
sharp cutoff at about $\ln(3)$, corresponding to a single particle
hopping back and forth across the boundary between the two halves of
the system. Both findings could be taken as an indication that there
is no particle 
%transport 
redistribution
deep in the MBL phase beyond the level of a
single particle and that the increase of the entanglement entropy is
solely due to configurational entanglement. The latter would also
imply that the asymptotic relation between entanglement and number
entropies derived in Ref.~\cite{KieferUnanyan1} for
non-interacting systems and demonstrated to hold also for interacting
particles in Ref.~\cite{KieferUnanyan2} of the form
\begin{equation}
\label{distr2}
S_N\sim \frac{1}{D^\nu}\ln S +\gamma,
\end{equation}
ceases to hold deep in the MBL phase. Here $\nu>0$ is an exponent of
the order of unity. In the following we argue that this interpretation
is too naive. The behavior of the probability distribution of entropies, observed in 
Ref.~\cite{LuitzBarLev}, is fully
consistent with the relation \eqref{distr2} as it implies a sub-exponential tail of $p(S_N)$.

%% JS
%First, as explained in the Suppl. Mat., we find from
%Eq.~\eqref{distr2} the following scaling of the variance of the number
%entropy
First, by assuming a small variation of the entanglement entropy from its
average value $\overline{S}$, i.e. $S= \overline{S}+\Delta S$ and
using Eq.~\eqref{distr2}, we find
\begin{eqnarray}
\overline{S}_N + \Delta S_N &\sim& \frac{1}{D^\nu} \Bigl(\ln(\overline{S}) +\ln\Bigl(1+ \frac{\Delta S}{\overline{S}}\Bigr)\Bigr) +\gamma \nonumber \\
&\approx& \frac{1}{D^\nu} \ln(\overline{S}) + \frac{1}{D^\nu} \frac{\Delta S}{\overline{S}} +\gamma.
\end{eqnarray}
From this we can read off the variance of the number entropy. Using, furthermore, the scaling of the
average entanglement entropy with system size $\overline{S}\sim L/D$ \cite{AbaninRev2019} we obtain
%
% \begin{eqnarray}
$ \Delta S_N \sim  {\Delta S}/{(D^\nu\overline{S})} \sim {\Delta S}/{L}$.
%
%Here we made use of the scaling of the disorder-averaged entanglement
%entropy $\overline{S}$ with system size $L$ in the MBL phase
%\cite{AbaninRev2019}. 
The variance of the number entropy therefore decreases with increasing
system size.
%and keeping in mind that eq.\eqref{distr2} holds in this form only for large values of $S$, 
There is thus no contradiction between the results in
Ref.~\cite{LuitzBarLev} and relation \eqref{distr2}.
 
\begin{figure}[h]
	\includegraphics*[width=0.98\columnwidth]{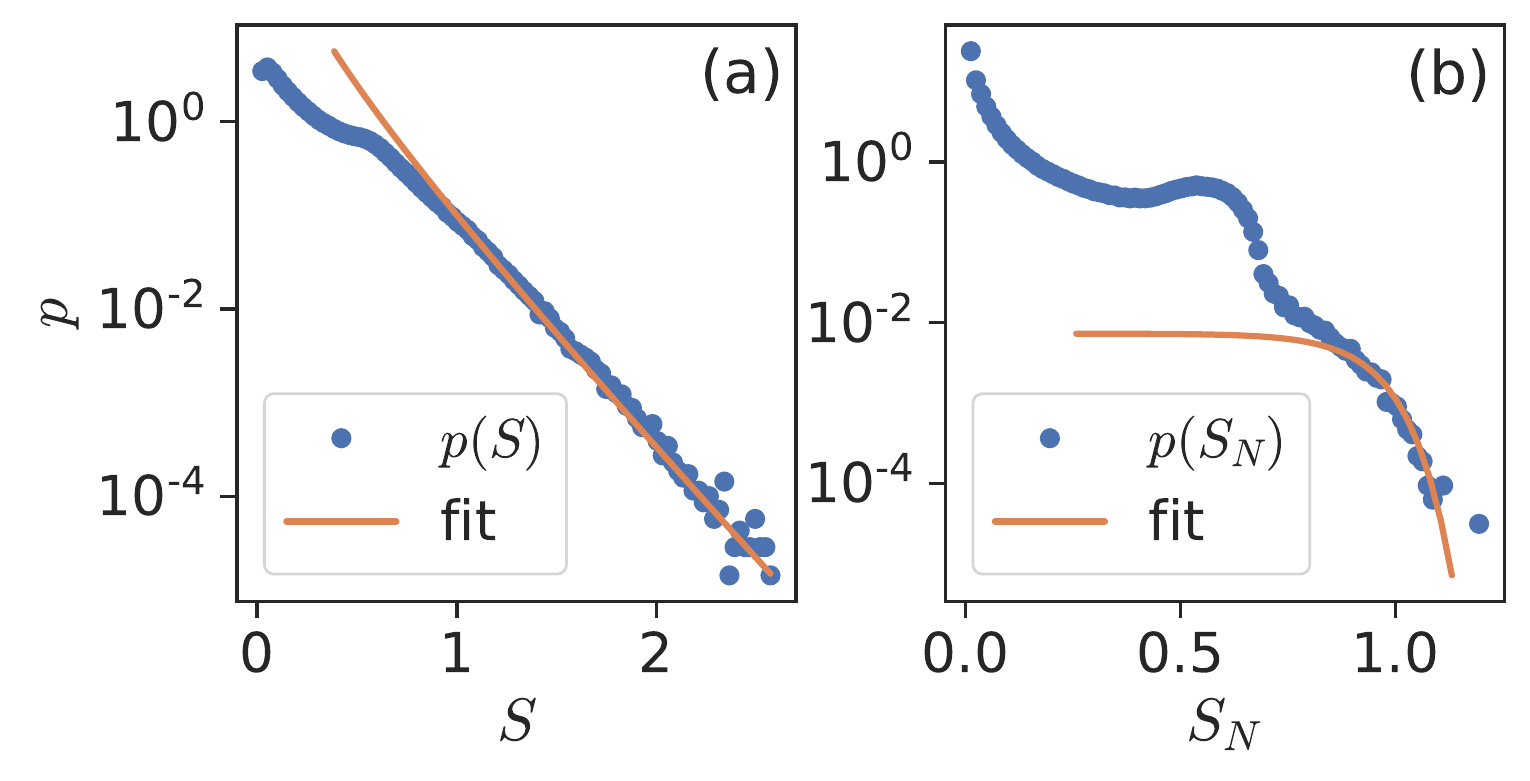}
	\caption{Distribution of saturation values for $D=40$ and
	$L=16$. (a) von-Neumann entropy $S$, and (b) number entropy
	$S_N$. Symbols denote numerical data from
	Ref.~\cite{LuitzBarLev}, orange lines are fits, see
	text.}  %If $p(S)$ has an exponential tail %(see
%	Eq.~\eqref{distr1}) and Eq.~\eqref{distr2} holds, then
%	$p(S_N)$ will drop rapidly beyond at a certain threshold
%	value.}
\label{Fig6f}
\end{figure}

The authors of Ref.~\cite{LuitzBarLev} found furthermore that
the probability distribution of the asymptotic entanglement entropy in
a finite system of length $L$ has an exponential tail, which we fit as
%
%\begin{equation}
%\label{distr1}
$p\bigl(S\bigr) \sim \exp\left(-{2 D S}/{L}\right)/S$.
%\end{equation}
%
This is shown in Fig.\ref{Fig6f}(a), where the distributions are based
on the data from Ref.~\cite{LuitzBarLev} and the fit is based on
the relation above. Note that the prefactor in the exponent is in
agreement with the asymptotic scaling $S \sim L/D$ found in
Ref.~\cite{AbaninRev2019}. If we plug Eq.~\eqref{distr2} into $p(S)$
we find, using $dS_N\sim dS/S$,
\begin{eqnarray}
 p\bigl(S_N\bigr)\sim  \exp\left\{-\frac{2D}{L}\exp\Bigl[\frac{D}{4}(S_N-\gamma)\Bigr]\right\} .\label{pSN}
\end{eqnarray}
This asymptotic expression shows a sharp cutoff for large values of
$D$ as soon as $S_N$ exceeds $\gamma$. Fig.~\ref{Fig6f}(b) shows a
comparison between the numerical data for $p(S_N)$ from
Ref.~\cite{LuitzBarLev} and the prediction \eqref{pSN} with
$\gamma$ used as a fitting parameter. The agreement is good. Thus the
seemingly sharp drop-off of the probability distribution $p(S_N)$
does not contradict the relation between number and entanglement
entropies found in Ref.~\cite{KieferUnanyan2} and is not a
sufficient indicator for a complete suppression of particle transport
beyond the level of a single particle.
%
%%%%%%%%%%%%%%%%%%%%%%%%%%%%%%%%%%%%%%%%%%%%
%%%%%%%%%%%%%%%%%%%%%%%%%%%%%%%%%%%%%%%%%%%%
%

Finally we note that the presence of a seemingly sharp drop in the
probability distribution $p(S_N)$ is consistent with the {\it absence
of localization} in other, exactly solvable models of non-interacting
fermions. The case of free fermions on a lattice with off-diagonal
(bond) disorder is discussed in App.~\ref{OffDiag}. This model is
known to be not fully localized, but the probability distributions of
the entropies show qualitatively the same behavior as for the
disordered Heisenberg chain.
%
% \begin{equation}
% \label{distr4}
% H=-\sum_j J_j (c_j^\dagger c_{j+1} + h.c.),
% \end{equation}
%
% where the random hopping amplitudes $J_j$ are drawn from a box
% distribution. For this model all properties can be calculated from the
% single particle correlation matrix allowing to study very large system
% sizes. It is known that the model is critical with a localization
% length that diverges at zero energy. This leads to the interesting
% scalings \cite{ZhaoSirker,KieferUnanyan1} $S \sim \ln\ln t\,$ and $S_N\sim\ln\ln\ln t $.
% Despite the fact that this model is known to be ergodic, the probability distributions 
% of the entanglement entropy and the number entropy show qualitatively the same
% behavior as that for the disordered Heisenberg chain found in 
% Ref.~[\cite{BarLevLuitz}]. In particular $p(S_N)$ shows a sharp
% drop-off. This is illustrated in Fig.~\ref{Fig7f} showing the entanglement distributions for this
% model.
%
%%%%%%%%%%%%%%%%%%%%%%%%%%%%%%%%%%%%%%%%
%\begin{figure}
%	\includegraphics*[width=0.99\columnwidth]{figure5_fu.pdf}
%	\caption{Distribution of saturation values for the free
%	fermion model with off-diagonal disorder (a) for the Von Neumann entropy and (b) for the number entropy, Eq.~\eqref{distr4}. }
%\label{Fig7f}
% \end{figure}
%%%%%%%%%%%%%%%%%%%%%%%%%%%%%%%%%%%%%%%%
%
We conclude that the features of the probability distributions for the
entanglement and number entropy found in Ref.~\cite{LuitzBarLev}
do not contradict the relation $S_N\sim \ln S$ and are therefore not
sufficient indicators for localization.

%%%%%%%%%%%%%%%%%%%%%%%%%%%%%%%%%%%%%%%%%%
\section{Hartley number entropy and number distribution} 
\label{Sec_Hartley}
% and probability flow to large number fluctuations}
%%%%%%%%%%%%%%%%%%%%%%%%%%%%%%%%%%%%%%%%%%
The pronounced drop-off of $p(S_N)$ shown in Fig.\ref{Fig6f} at
$S_N\sim \ln 3\approx 1.098$ could be taken as indication that at
sufficiently large times only a single particle fluctuates between the
two halves of the system. If the system is localized, one expects that
the probability distribution of particle numbers $p(n)$ in one
partition for a given realization and initial state develops a sharp
maximum at some value $n_\textrm{max}$ in the thermodynamic limit
after a transient. Thus allowing fluctuations of a single particle one
would expect non-vanishing probabilities only for the three values of
$n$, $n=n_\textrm{max}$, and $n=n_\textrm{max}\pm 1$, limiting the
number entropy to values less than $\ln 3$. Since the number entropy
does not exceed $\ln 3$ in our simulations at strong disorder, a
possible scenario consistent with the standard picture of MBL would be
a long transient redistribution of probabilities within the restricted
range $n_\textrm{max}\pm 1$.

%%%%%%%%%%%%%%%%%%%%%%%%%%%%%%%%%%%%%%%%%%
\begin{figure}[h]
	\includegraphics*[width=0.9\columnwidth]{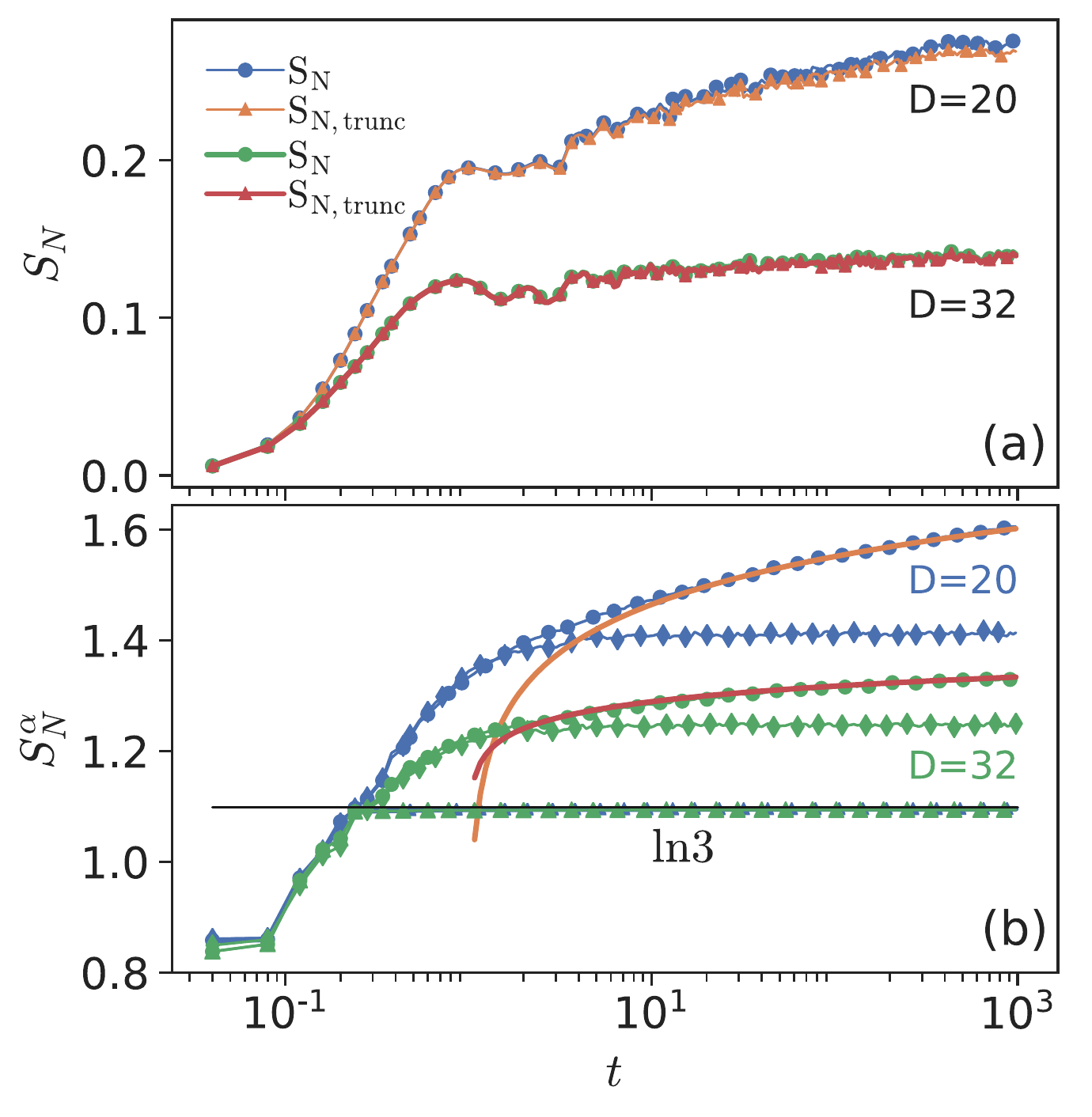}
	\caption{(a) $S_N$ for the full and the truncated distribution
	$p(n)$, where only contributions from $p(n_\textrm{max}),
	p(n_\textrm{max}\pm 1)$ are taken into account. (b) R\'enyi
	number entropies $S_N^{(\alpha)}$ for $\alpha = 0.001$
	(circles) and double logarithmic fits (lines). Also shown is
	the Anderson case (diamonds), i.e.~Eq.~\eqref{tV} with
	$V=0$. $p(n)$ is truncated at $p_c= 10^{-10}$. The full
	entropies are compared to those where only $p(n_\textrm{max})$
	and $p(n_\textrm{max}\pm 1)$ are taken into account. The
	latter approach the maximum value of $\ln 3$.}
\label{FigRenyi}
\end{figure}
%%%%%%%%%%%%%%%%%%%%%%%%%%%%%%%%%%%%%%%%%%

To assess the possibility of such a strictly bounded redistribution of
probabilities, we have calculated the time evolution of $S_N$ from a
truncated distribution taking into account only the values
$p(n_\textrm{max})$ and $ p(n_\textrm{max}\pm 1)$.
Fig.~\ref{FigRenyi}(a) shows a comparison of the full with the
truncated number entropy for two different disorder strengths. One
recognizes---in particular for the larger disorder value---that $S_N$
in this regime is indeed dominated by those three probabilities. The
number entropy is, however, insensitive to the dynamics in the tails
of the probability distribution. Due to the extremely slow growth of
number fluctuations, reflected in the $\ln\ln t$ scaling of the number
entropy, the probabilities for large number fluctuations will remain
very small for numerically accessible time-scales. Nevertheless, these
fluctuations will eventually become large and destroy localization if
they continue to grow. It is thus important to consider a quantity
that is sensitive also to the tails of the number distributions. 
A potential candidate for such
a quantity is the Hartley number entropy which is the R\'enyi entropy,
Eq.~\eqref{SN_alpha}, of degree $\alpha= 0$. The Hartley entropy is
the logarithm of the cardinality of $p(n)$, i.e.~it counts the number
of configurations with probabilities different from zero. 

%% JS 
Since quantum mechanically the unitary time evolution immediately
leads to a non-zero probability for any particle distribution
(although most of them will be extremely small) consistent with total
particle number conservation independent of whether or not the system
is localized, it is important to introduce a cutoff $p_c$ and to only
consider configurations with $p(n,t)>p_c$. All values below the cutoff
are set to zero and the distribution is renormalized. The important
point then is that for a localized system this truncated Hartley number entropy with
{\it any} cutoff $p_c>0$ has to saturate in the thermodynamic limit, i.e.,
there can only be a finite number of configurations with $p(n)>p_c$
for long times. The saturation value will, of course, depend on the
cutoff $p_c$.

\begin{figure}[h]
	\includegraphics*[width=0.99\columnwidth]{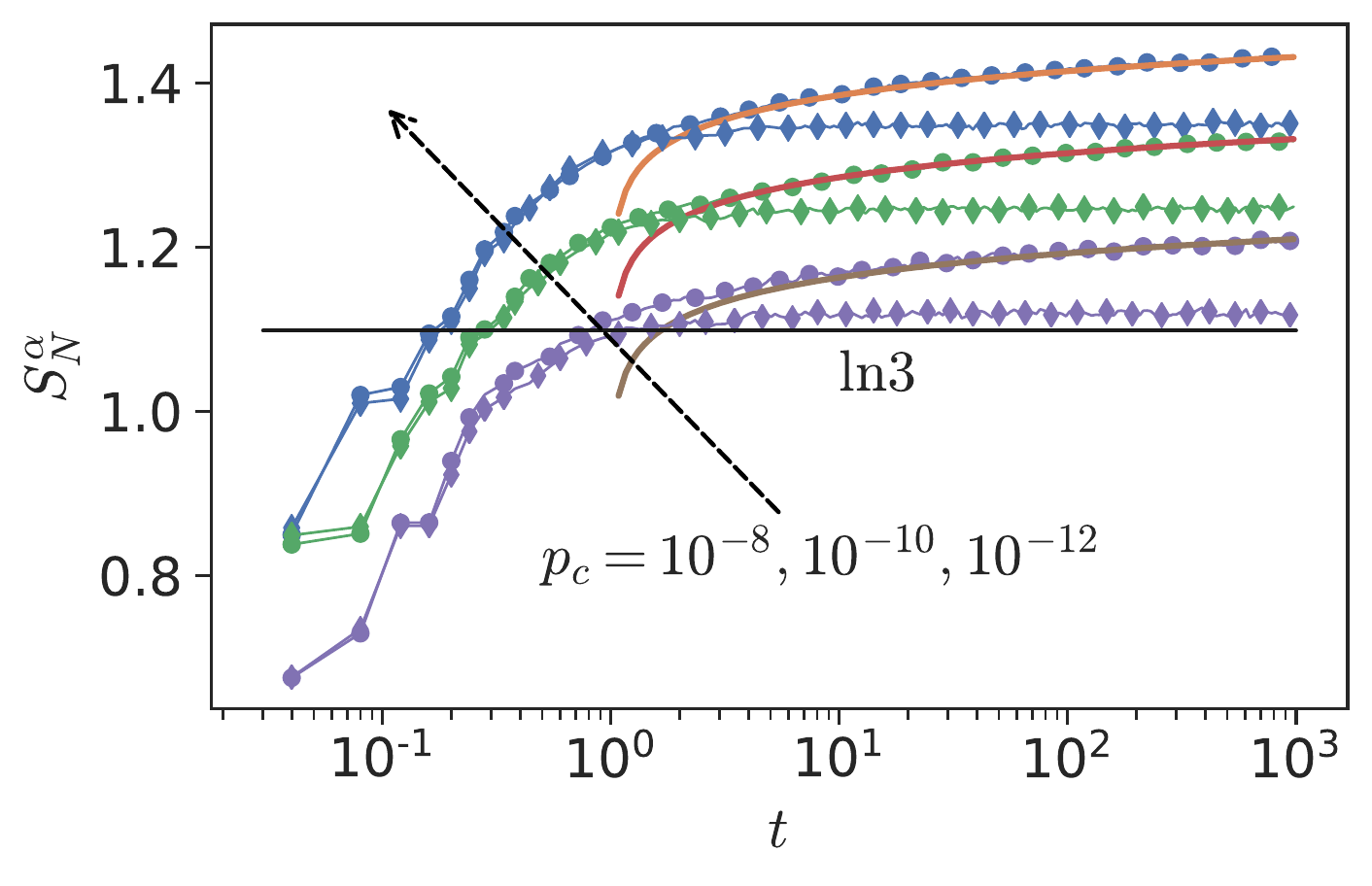}
	\caption{R\'enyi number entropy $S_N^{(\alpha)}(t)$ for
	$\alpha=0.001$, $D=32$, and different values of
	$p_c$. $S_N^{(\alpha)}$ increases with decreasing cutoff value
	for the MBL case, leading effectively to a simple constant
	shift. For the Anderson case, on the other hand,
	$S_N^{(\alpha)}(t)$ always saturates.}
\label{Fig8f}
\end{figure}

We here choose a very small but non-vanishing value of $\alpha$ and
calculate the time evolution of $S_N^{(0.001)}$. The results for the
Hartley entropy are shown in Fig.~\ref{FigRenyi}(b) for
$p_c=10^{-10}$. For each disorder realization, $p(n,t=0)=1$ for $n$
corresponding to the initial number of particles in the partition and
zero otherwise. The truncated Hartley entropy for each realization---and
consequently also the average---is therefore zero at $t=0$. The
entropy then continues to increase $\sim\ln\ln t$ well above the value
of $\ln 3$. Even more importantly, we do not find any signatures for a
saturation for all numerically accessible times. Also shown is the
result for the Anderson case, i.e.~Eq.~\eqref{tV} with $V=0$. Here the
Hartley entropy saturates, which is consistent with a strict
localization of particles. Fig.~\ref{Fig8f} shows that while the values
of the entropies for the MBL and Anderson case do depend on the chosen
cutoff $p_c$, the qualitative behavior is independent of $p_c$.
%%%%%%%%%%%%%%%%%%%%%%%%%%%%%%%%%%%%%%%%%%%%
%%%%%%%%%%%%%%%%%%%%%%%%%%%%%%%%%%%%%%%%%%%%

We can define the occupied particle number state $\tilde n(p_c)$ which
is furthest away from the most likely value $n_\textrm{max}$ while
still obeying $p(\tilde n)\ge p_c$. The dynamical behavior of the
truncated Hartley number entropy, shown in Fig.~\ref{FigRenyi}, must then be
understood as an increase of $\tilde n$ according to
\begin{equation}
\tilde n\,  \sim \, (\ln t)^\beta,
\end{equation} 
where $\beta \leq \frac{1}{2}$.
In other words, the width of $p(n)$ measured at $p_c$ is increasing
logarithmically in time. This must be interpreted as a constant flow
of probability to higher particle number fluctuations.

%%%%%%%%%%%%%%%%%%%%%%%%%%%%%%%%%%%%%%%%%%
\section{Conclusions}
\label{Concl}
%%%%%%%%%%%%%%%%%%%%%%%%%%%%%%%%%%%%%%%%%%
We have presented a detailed study of particle number fluctuations in
the putative many-body localized (MBL) phase of the isotropic
Heisenberg model. Our results point to an absence of true
localization; particles continue to spread through the system at a
very slow rate even for strong disorder, far from the ergodic-MBL
transition. Our conclusions are based on two main findings: (1) For
all disorder strengths $D>D_c$ investigated, the time regime where
$S\sim\ln t$ holds in a finite system is exactly the same where
$S_N\sim\ln\ln t$ holds. A saturation of $S_N(t)$ while $S(t)$
continues to grow is never observed. 
%% JS
We have also shown that the growth of $S_N$ is not a consequence of
rare regions but rather represents typical behavior. (2) For all
disorder strengths $D>D_c$ investigated, the Hartley number entropy
grows as $S_N^{(\alpha\to 0)}\sim\ln\ln t$ and reaches value larger
than $\ln 3$. The width of the distribution $p(n)$ measured at some
small cutoff $p_c$ thus grows $\sim\ln t$: There is a constant flow
towards higher particle number fluctuations.

In addition, we have also shown that the sharp cutoff in the
distributions of number entropies at $S_N\sim\ln 3$ observed in
Ref.~\cite{LuitzBarLev} does not contradict the relation
$S\sim\exp(S_N)$ established in
Refs.~\cite{KieferUnanyan1,KieferUnanyan2} but is rather fully
consistent with it. Other arguments in favour of a full localization
given in
Ref.~\cite{LuitzBarLev} were based on a study of the saturation values of $S_N$. 
This quantity, however, is difficult to analyze because the long
saturation times for large systems are causing numerical issues, the lack
of a known scaling, and the possible non-monotonicity of the
saturation values as function of system size. These issues are
discussed further in App.~\ref{Sec_Sat}.

%\clearpage

\acknowledgments
J.S. acknowledges support by the Natural Sciences and Engineering
Research Council (NSERC, Canada) and by the Deutsche
Forschungsgemeinschaft (DFG) via Research Unit FOR 2316. We thank
D.~Luitz for discussions and are grateful for the computing resources
and support provided by Compute Canada and Westgrid. M.K., R.U. and
M.F. acknowledge financial support from the Deutsche
Forschungsgemeinschaft (DFG) via SFB TR 185, project number
277625399. The simulations were (partly) executed on the high
performance cluster "Elwetritsch" at the University of Kaiserslautern
which is part of the "Alliance of High Performance Computing
Rheinland-Pfalz" (AHRP). We kindly acknowledge the support of the
RHRK.

\appendix
%%%%%%%%%%%%%%%%%%%%%%%%%%%%%%%%%%%%%%%%%%%%%%%
\section{$p(S_N)$ for free fermions with off-diagonal disorder}
\label{OffDiag}
%%%%%%%%%%%%%%%%%%%%%%%%%%%%%%%%%%%%%%%%%%%%%%%

\begin{figure}[h]
	\includegraphics*[width=0.99\columnwidth]{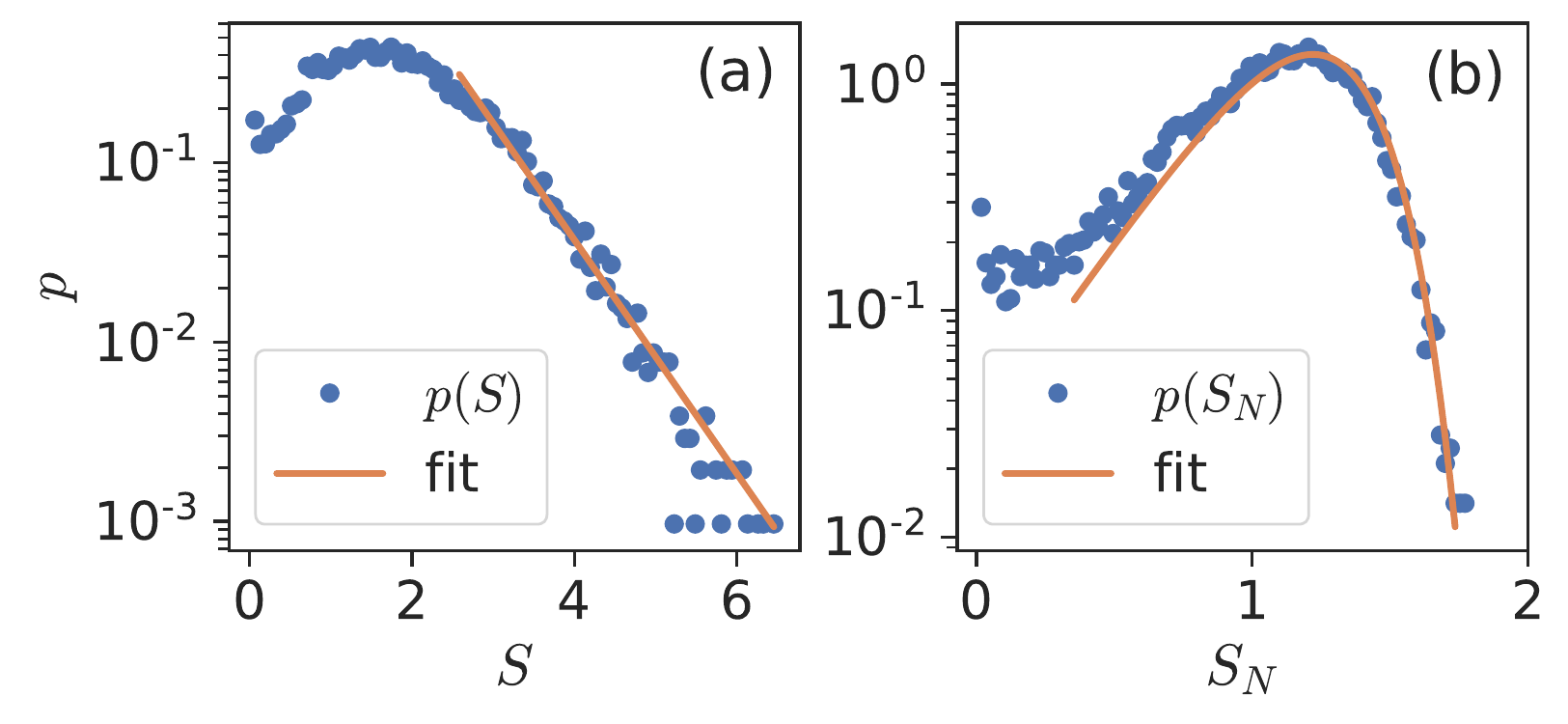}
	\caption{Distribution of saturation values for the free
	fermion model with off-diagonal disorder, Eq.~\eqref{distr4},
	for $L=1024$ using $20,000$ disorder realizations:  (a)
	von-Neumann entropy, and (b) number entropy. Symbols
	represent numerical data, the line in (a) is an exponential
	fit $p(S)\sim \mathrm{e}^{-\frac{3}{2}S}$, and the line in (b)
	the corresponding fit for $S_N$ using the relation \eqref{distr2},
	$p(S_N)\sim \mathrm{exp}[4(S_N-\gamma)] \mathrm{exp}\{
	-\frac{3}{2}\mathrm{exp}[4(S_N-\gamma]\}$ with $\gamma=\ln
	3.75$. Here the constants are fit parameters.}
\label{Fig7f}
\end{figure}

Here we want to demonstrate that also for exactly solvable models which are known to {\it not} fully localize,
the probability distribution of the number entropy has a seemingly sharp decline.
To this end, we consider free fermions on a lattice with off-diagonal (bond) disorder
\begin{equation}
\label{distr4}
H=-\sum_j J_j (c_j^\dagger c_{j+1} + h.c.),
\end{equation}
where the random hopping amplitudes $J_j$ are drawn from a box
distribution. For this model all properties can be calculated from the
single particle correlation matrix allowing to study very large system
sizes. It is known that the model is critical with a localization
length that diverges at zero energy. This leads to the interesting
scalings \cite{ZhaoAndraschkoSirker,KieferUnanyan1} $S \sim \ln\ln t\,$ and $S_N\sim\ln\ln\ln t $.
Despite the fact that this model is known to not fully localize, the probability distributions 
of the entanglement entropy and the number entropy show qualitatively the same
behavior as for the disordered Heisenberg chain found in 
Ref.~\cite{LuitzBarLev}. In particular, $p(S_N)$ has a sharp
drop-off. This is illustrated in Fig.~\ref{Fig7f}. 
%
%%%%%%%%%%%%%%%%%%%%%%%%%%%%%%%%%%%%%%%%
%%%%%%%%%%%%%%%%%%%%%%%%%%%%%%%%%%%%%%%%
%

%%%%%%%%%%%%%%%%%%%%%%%%%%%%%%%%%%%%%%%%%%%%%%%
\section{Saturation values of entropies}
\label{Sec_Sat}
%%%%%%%%%%%%%%%%%%%%%%%%%%%%%%%%%%%%%%%%%%%%%%%

Here we want to address the question of what information can be
obtained from trying to extrapolate the saturation values of the
number entropy in system size. We will demonstrate that no clear
scaling law emerges from numerical simulations for the available
system sizes. Note that the scaling function is not known a priori.

\begin{figure}[h]
	\includegraphics*[width=0.99\columnwidth]{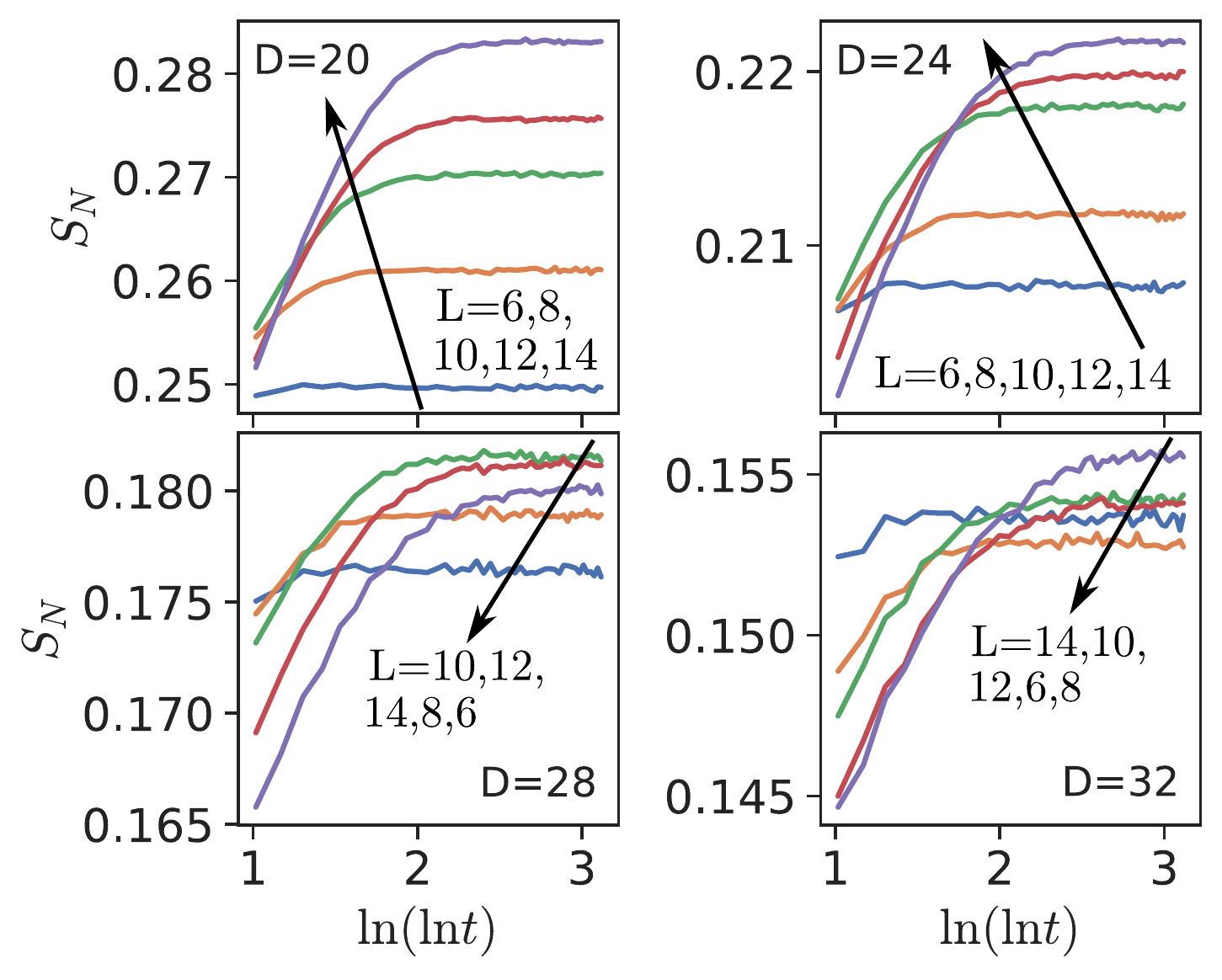}
	\caption{$S_N(t)$ as a function of $\ln\ln t$ for
	$L=6,8,10,12,14$ and different $D$. The saturation values show
	no clear, consistent scaling with system size. Note, in
	particular, that the saturation values scale non-monotonically
	with $L$ for $D=24,28,32$.}
\label{Fig4f}
\end{figure}

We start by showing in Fig.~\ref{Fig4f} the results for $S_N(t)$ for
four different disorder strengths and various system sizes.

%%%%%%%%%%%%%%%%%%%%%%%%%%%%%%%%%%%%%%%%%%%
%%%%%%%%%%%%%%%%%%%%%%%%%%%%%%%%%%%%%%%%%%%
By comparing the different disorder strengths, it is obvious that
there is no simple scaling function $f(D,L)$ of the saturation values
as function of system size $L$ and disorder strength $D$. Secondly,
the scaling in system size is not monotonic for $D=24,28,32$. I.e.,
the saturation value as function of system size can show a 'dip' which
is not indicative of the thermodynamic limit, making any extrapolation
difficult. Finally, we note that the saturation times are roughly
increasing exponentially with system size and already reach times
$\sim 10^9$ for $L=14$. Since the calculations are performed in double
precision, times $t\gtrsim 10^{14}$ are not accessible and averaging
over times beyond what is reliably possible in double precision can
potentially lead to incorrect results. Overall, it appears to be very
difficult to make any reliable statements about the scaling of the
saturation values of the number entropy based on exact
diagonalizations of small systems in double precision at very large
disorder.

\begin{figure}[htb]
	\vspace{3mm}
	\includegraphics*[width=0.9\columnwidth]{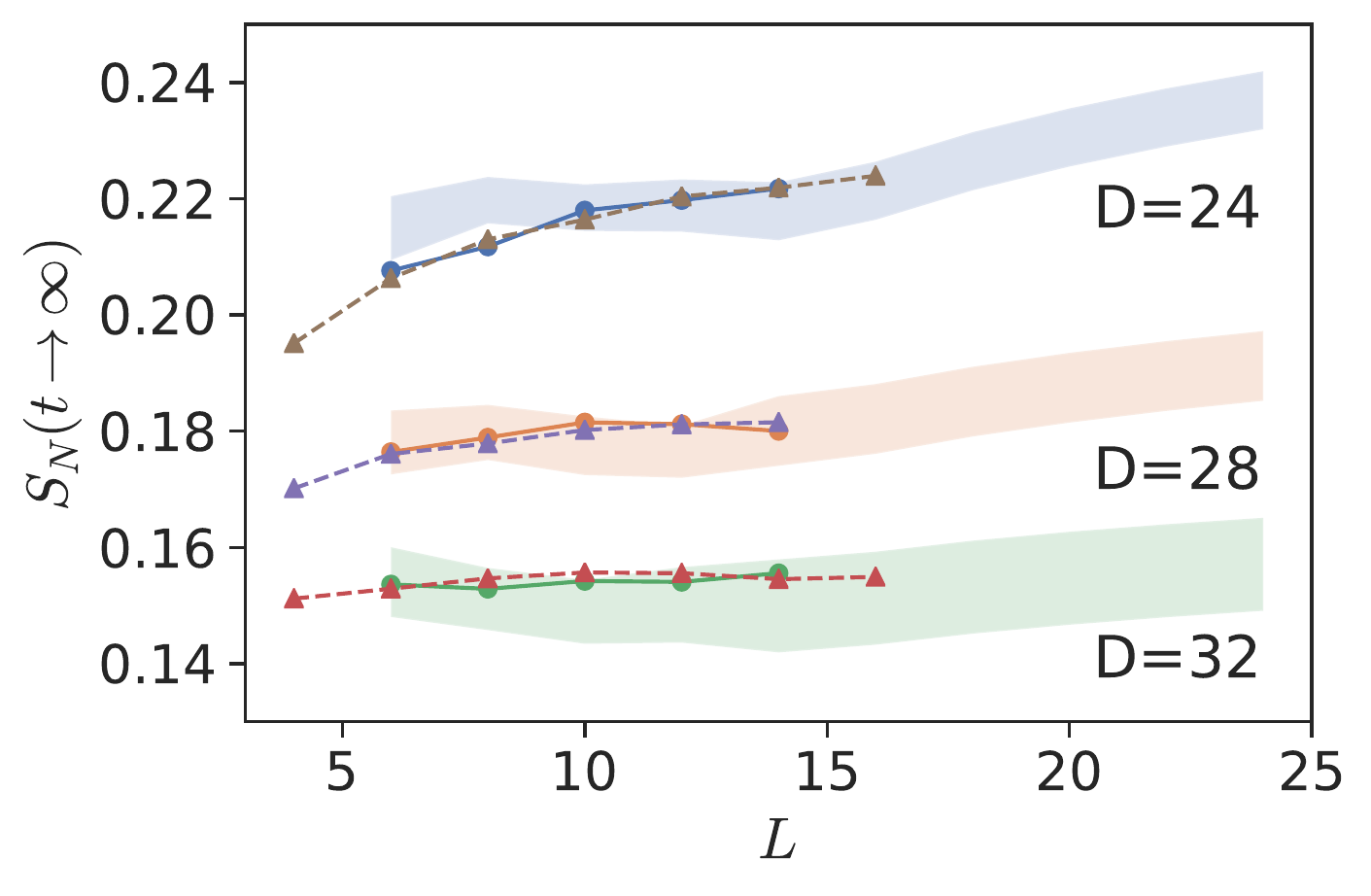}
	\caption{Estimates for the saturation values
	$S_N(t\to\infty)$. For system sizes up to $L=14$, the
	saturation values are obtained directly from exact
	diagonalizations (full circles) and compared to the values
	from Ref.\cite{LuitzBarLev} (triangles) for disorder
	strengths $D=24, 28$ and $32$. For larger system sizes the
	method described in the text is used, leading to the shaded
	bands.}
\label{Fig5f}
\end{figure}

Our best try to estimate the saturation values for system sizes up to
$L=24$ is shown in Fig.~\ref{Fig5f}. 
%%%%%%%%%%%%%%%%%%%%%%%%%%%%%%%%%%%%%%%%%%%%
%%%%%%%%%%%%%%%%%%%%%%%%%%%%%%%%%%%%%%%%%%%%
For system sizes $L>14$ we proceed as follows: First, we extrapolate
the saturation times $t_{\textrm{sat}}\sim\exp(L)$, obtained for
smaller system sizes, in $L$, see Suppl.~Mat.~of
Ref.~\cite{KieferUnanyan2}. Second, the double logarithmic fit
of $S_N$ obtained for smaller times is used to determine the
saturation value, $S_N(t\rightarrow\infty) \approx (\nu/2)\ln\ln
t_{\textrm{sat}} +b$. This leads to the shaded bands with the width of
the shaded bands being a consequence of the uncertainty in estimating
$t_{\textrm{sat}},\nu$ and $b$. As we have already seen in
Fig.~\ref{Fig4f} for smaller system sizes, the scaling of the
saturation value with $L$ is, in general, non-monotonic. We note, in
particular, that also for $D=32$ the saturation value appears to {\it
increase} for system sizes $L\gtrsim 16$. We thus believe that the
interpretation in Ref.~\cite{LuitzBarLev} of the decrease of the
saturation values in a certain range of system sizes as an indication
of a saturation in the thermodynamic limit is not justified.

A more useful approach---less prone to issues with the finite-size
scaling---is to study the dependence of the time scale where $S$
($S_N$) start to deviate from a logarithmic (log-log) scaling. This
point has been investigated in Sec. II.A of the main text and the
corresponding time scales are indicated in Fig.~2 in the same section.

%\bibliography{Literatur.bib}
%\bibliography{/Users/js/Literatur.bib}

\end{document}